# Characterizing and Evaluation :Temporal properties of real and synthetic datasets for DTN

Hemal Shah
Ganpat University
North Gujarat, India
hemal.shah@ganpatuniversity.ac.in

Yogeshwar Kosta
Marwadi Education Foundation
Rajkot, Gujarat, India
ypkosta@gmail.com

Vikrant Patel
Ganpat University
North Gujarat, India
research.vikrantpatel@gmail.com

*Abstract*— Node's movements play a significant role in disseminating messages in the sparse mobile adhoc network. In the network scenarios, where traditional end-to-end paths do not exist, mobility creates opportunities for nodes to connect and communicate when they are encountered. A series of encountering opportunities spread a message among many nodes and eventually deliver to the destination. Further improvements to the performance of message delivery can come from exploiting temporal mobility properties. It is modeled as time varying graph, where, moving nodes are considered as vertices and contact opportunity to other nodes as an edge. The paper discusses about characterization and design of the temporal algorithm. Then, evaluating temporal distance, diameter and centrality of real and synthetic mobility data sets.

**Introduction**

There are situations in mobile ad-hoc networks, where nodes are completely disconnected and may rely on relay nodes for contact opportunities to transfer the message. Such relay nodes create an opportunity for partial connectivity and carry the message until the next node or destination comes into contact[1]. In other networks, connectivity may exist, but only occasionally or intermittently. This intermittent connectivity is not failure or fault but, rather an integral part of dynamic networks. These networks are called delay tolerant networks (DTNs)[2]. DTN utilizes a Store-Carry-Forward[3] mechanism in which the intermediate node stores messages and forwards them to nodes it encounters. In this manner, messages could be delivered to the destination hop-by-hop even if no stable end-to-end path exists. As network partition occurs frequently, if only one replica of the message[4] is kept, the message may reach the edge of partitioned network and be failed to be delivered at the destination. In order to increase the message delivery rate, each node can keep forwarded messages and copy them to other nodes it encounters. In this multiple-copy routing[5] manner, several replicas of the same message exist within the network.

In both single-copy routing and multiple-copy routing, the message delivery rate depends upon the node mobility, network connectivity and the intermediate node chosen strategy. Based on sufficient network connectivity, message delivery strategies should utilize node's mobility characteristics to increase the message delivery rate and reduce network overhead. Classic studies looked[6]at analyzing static or aggregated networks, i.e., networks that do not change over time or built as the results of aggregation of information over a certain period of time. Given the soaring collections of measurements related to very large, real network traces, researchers[7] are quickly starting to realize that connections are inherently varying over time and exhibit more dimensionality than static analysis can capture.

In time varying graph or temporal graph[6][8] vertices represent the node and opportunistic contact between nodes represent edge or links. This

links are changing over the time and raising interesting questions:

- Are there any metrics [9][10][11] evolved or proposed by researchers relating to temporal graphs in DTN?
- If available, can they be used to analyze real and synthetic data sets?
- Can time varying behavior of mobile ad-hoc network be used for designing DTN routing algorithm?

This has motivated to contribute towards defining the metrics related to temporal graph to DTN. Then, designing the temporal algorithm to evaluate metrics from real trace and synthetic data sets. Author's contributions are:

1. Modeling DTN as time varying graph.
2. Defining temporal measurement: temporal distance, centrality, betweenness, diameter.
3. Design and implementation temporal measurement algorithm.
4. Evaluation of real and synthetic trace datasets for temporal properties.

Section II discusses DTN as time varying graph and defining temporal indicators Section III discusses design temporal algorithm Section IV presents real and synthetic dataset evaluation for temporal properties.

I. DTN AS TEMPORAL GRAPH AND DEFINING RELATED METRICS

In static or aggregated networks, it is observed [10] that the connections ( in case of ad-hoc network/mobile ad-hoc network/DTN ) are inherently varying over time and exhibit more dimensionality[12] than static analysis can capture. Static graphs treat all links as appearing at the same time. It is unable to capture key temporal characteristics, and gives an overestimate of potential paths, connection pairs of nodes which cannot provide any information on the delay associated with information spreading process. Thus, to represent DTN as temporal graphs, the mobile nodes can be presented as vertices and opportunistic contact between nodes as an edge. In understanding duration of contact, inter-contact time, repeated contact, time order of contact along a path on time interval basis, Temporal graph[6] is represented by sequence of time windows, for each window is considered a snapshot of the network at that time interval. The temporal distance, diameter and centrality metrics evolves over this view of the temporal graph retain the time ordering, repeated occurrences of connections between nodes, contact time and deletions of edges.

**Temporal Graph:**

Given a network trace starting at $T_{min}$ and ending at $T_{max}$, a contact between nodes, $i,j$ at time 's' is defined with the notation $R_{ij}^s$. A temporal graph[6] $\mathcal{G}_t^\omega(T_{min}, T_{max})$ with $N$ nodes consists of a sequence of graphs $G_{t_{min}}$, $G_{T_{min}+w}$,…, $G_{T_{max}}$, where '$w$' is the size of each time window unit e.g., seconds. Then, $G_t$ consists of a set of nodes $V$ and a set of edges $E$ such that $i,j \in V$, if and only if, there exists $R_{ij}^s$ with $t \leq s \leq t + w$.

**Temporal Path:**

For given two nodes $i$ and $j$ temporal path defines as:

$p_{ij}^h(T_{min}, T_{max})$ To be the set of paths starting from $i$ and finishing at $j$ that passes through the nodes $n_1^{t_1} \ldots n_i^{t_i}$, where $t_{i-1} \leq t_i$ and $T_{min} \leq t_i \leq T_{max}$ is the time window, that node $n$ is visited and $h$ is the max hops within the same window $t$. There may be more than one shortest path.

**Temporal Distance:**

Given two nodes $i$ and $j$, the shortest temporal distance defines as:

$$d_{ij}^h(T_{min}, T_{max})$$

to be the shortest temporal path length, starting from time $T_{min}$, this can be thought as the number of time windows (or temporal hops) which takes for information to spread from a node $i$ to node $j$. The horizon $h$ indicates the maximum number of nodes within each window $G_T$ through which information can be exchanged, or in practical terms, the speed that a message travels. In the case of temporally disconnected node pairs $q,p$ i.e., information from $q$ never reaches $p$, then set the temporal distance $d_{pq} = \infty$.

**Temporal Centrality**

Finding out the most central node is important, because it helps disseminating information faster, stops epidemics and protects the network from breaking. It is also known as 'important node' or 'special node' as it quickly spreads the information to many nodes mediating between the most information flows. The Figure 1 shows the different meanings of centrality[13][14] as under:

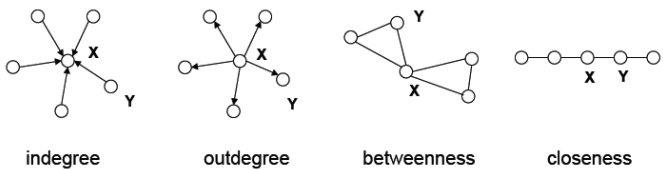

Figure 1 Different meanings of Centrality

**Closeness Centrality**

$C_i^d$ = number of links to i (indicates the popular node) Where, $C_i = \sum_{i \ne j} d_{ij}$ Average shortest path length to all other nodes E.g. In figure 2, closeness centrality [13]of central node is 6 (highest).

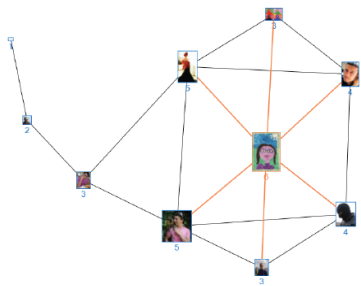

Figure 2    Node Degree

**Betweenness Centrality**

It defines number of shortest path passing through node.

$$C_i^{bet} = \sum_{i \ne s \ne t} \delta_s t(i) / \delta_{st}$$

Where, $\delta_{st}$ is number of shortest paths from s to t and $\delta_s t(i)$ is number of shortest paths passing through i

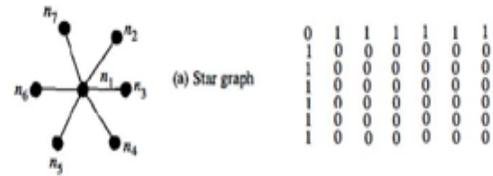

Figure 3    Betweenness centrality

In Figure 3 (star graph), shortest distance between all the nodes s and d is passing from node n1. So, number of total shortest path between s and d = (6*6) = 36. Number of total shortest path between s and d passing from n1 = 36. Therefore, $C_{n1}^{bet} = \frac{36}{36} = 1.0\ i.e,$ betweenness centrality[9] of node n1 in star diagram is 1.0, which is highest among all the nodes.

**Temporal Closeness Centrality**

It defines average shortest temporal paths to all other nodes. Given the shortest temporal distance dij(T$_{min}$; T$_{max}$), the temporal closeness centrality[14] C is expressed as:

$$C_i^h = \frac{1}{W(N-1)} \sum_{j \ne i \in V} d_{i,j}^h$$

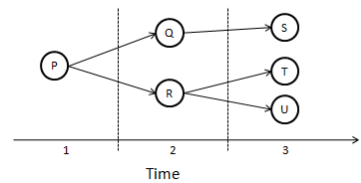

Figure 4  Temporal closeness centrality

As shown in Figure 4, temporal closeness centrality of node p at time interval is:

$$C_P = \frac{(2+2)+(3+3+3)}{(3*(6-1))} = 0.867$$

## Temporal Betweness Centrality

The temporal betweenness centrality of node i has been defined as the fraction of temporal shortest paths that passes through i. For an example, let us consider the case where nodes i and j being connected by just one shortest path pij = (it0; kt1 ; jt2 ), so that a message from i to j has to pass through k first at time t1 before being delivered to j at time t2. Since the path through k is the only way for i to send a message to j, then k plays an important role of mediatory and thus plays a 'central' for communication between i and j.

$$C_i^B(t) = \frac{1}{(N-1)(N-2)} \sum_{\substack{j \leq V \\ j \neq i}} \sum_{\substack{k \in V \\ k \neq i, k \neq j}} \frac{U(i,t,j,k)}{|S_{jk}^h|}$$

Where,
U(i, t, j, k) returns number of shortest paths from j to k, which node i is holding a message at at time window t , $|S_{jk}^h|$ indicates number of shortest temporal paths between j and k

Here, sum over all time windows for each node.

$$C_i^B(t) = \frac{1}{W} \sum_{t=1}^{W} C_i^B \ ((t \times w) + t_{min})$$

## Temporal Diameter

The Temporal Diameter [12] is defined as the maximum distance in the system, taken over all pairs of nodes. It can also be defined as the maximum of all eccentricities in the system.

## III    Temporal Algorithm

Temporal distance $d_{ij}(T_{min}, T_{max})$, is computed in terms of number of time windows i.e. $d_{ij}(T_{min}, T_{max}) = d_{ij}^t(T_{min}, T_{max})$. Next, algorithmic steps are described to compute $d_{ij}(T_{min}, T_{max})$.

For each pair of i and j, algorithm computes $d_{ij}(T_{min}, T_{max})$ and then, takes average of all values. This way temporal distance is computed in number of time stamps. If average value multiplies with w, then result is the temporal distance in terms of time (in seconds). Eq.(1) gives average temporal distance between $T_{min}$ and $T_{max}$.

$$L(T_{min}, T_{max}) = \frac{\omega}{N(N-1)} \Sigma_{ij} \ d_{ij} \ (T_{min}, T_{max}) \quad \ldots 1$$

## Timewindow (w) Calculation

To understand the computation of Time window, refer Table 1 below showing calculation on dataset as an example, where each cell value represents total contact time between a particular pair i,j divided by total number of contact occurrences. For each node pair (i,j) compute a sum of all values. It returns average meeting time per contact. The optimal value of time window is greater than average meeting time, because if time window<= average meeting time, then in most of the time windows, number of contact occurrence will be around one. That means, information cannot be diffused efficiently into the network.

Table 1        Time window Calculation

| Node ID | 1 | 2 | 3 | 4 | 5 | 6 | $\Sigma \frac{\text{Total Contact}}{\text{Total No. of Oc}}$ |
|---|---|---|---|---|---|---|---|
| 1 | 0/0 | 480/2 | 720/3 | 480/2 | 960/4 | 480/2 | 3120/13 |
| 2 | 500/2 | 0/0 | 750/3 | 250/1 | 1000/4 | 500/2 | 3000/12 |
| 3 | 735/3 | 490/2 | 0/0 | 490/2 | 735/3 | 245/1 | 2695/11 |
| 4 | 235/1 | 940/4 | 1175/5 | 0/0 | 705/3 | 470/2 | 3525/15 |
| 5 | 1300/5 | 260/1 | 780/3 | 520/2 | 0/0 | 1040/4 | 3900/15 |
| 6 | 510/2 | 510/2 | 255/1 | 1530/6 | 1275/5 | 0/0 | 4080/16 |
| $\Sigma T_{ij}(T_{min}, T_{max})$ | | | | | | | 20320 |
| $\Sigma N_{ij}$ | | | | | | | 82 |
| ANS | | | | | | | 247.80 |

From above computation it is established that, for effective information diffusion process into the network optimal time window should be greater than $\frac{\Sigma T_{ij}(T_{min},T_{max})}{\Sigma N_{ij}}$ i.e., time window = 300 seconds, resulting into total number of time window = $(T_{max} - T_{min})/w = (900 - 0)/300 = 3$ timestamps.

**Computation of Temporal Distance**

For each pair of nodes, i,j from graph G, let's find $d_{ij}^t(t_0, t_{900})$. Before starting calculation of temporal distance of each pair i,j, initialize number of empty lists equal to that of calculated number of time window.

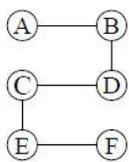
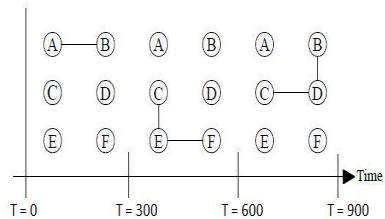

Figure 5 Scenario of aggregate graph for hiding time varying properties

Figure 6 Scenario of temporal graph for evaluating temporal distance

For each pair (i, j), i≠j, start scanning timestamps from 1 to 3. For each timestamp, add occurred node id into respective list of timestamp.

If node A wanted a piece of information to reach F, according to the static graph in Figure 5, it could do so via nodes B, C, D, and E. Also, reversing the path, if node F wanted to reach A, it could do so i.e., suggesting paths are symmetric. In fact, contacts between A and F occur in the wrong time order to facilitate this. The static graph incorrectly shows that the information could spread between node A and node F. Next, it is shown how temporal algorithm calculates the temporal distance between nodes. Here, $T_{min} = 0$ and $T_{max} = 900$.

**Pre-conditions**

Pair of node (i,j) occurs whenever there is an contact edge between node pair (i,j).

**Case 1**
If I == j then, return 0, in computed matrix below, temporal distance (A,A) = (B,B) = (C,C) = (D,D) = 0

**Case 2**
If both i and j occurs in same timestamp then return ($j^{th}$ timestamp number – $i^{th}$ timestamp number) or return (0). In Figure 6, node A and node B occurs in same timestamp no. 1 , so temporal distance between A and B is (B's timestamp no. – A's timestamp no.) = (1-1) = 0 timestamp.

**Case 3**
If i occurs earlier than j, then search occurrences of j in consecutive timestamps by using other occurred nodes in same timestamp in which i has occurred; for each pair i,j it may give more than one path in terms of required timestamp, in such a case select the shortest timestamp.

In Figure 6, for temporal distance (A,D), node A occurred at timestamp number 1 and node D occurred at timestamp number 3. Also, there is an intermediate node B which is common between node A and node D. So temporal distance (A,D) = (node D's timestamp number – node A's timestamp number) = (3 – 1) = 2 timestamps

**Case 4**
If i occurs and j does not occur during consecutive timestamp till $T_{max}$, then temporal path between pair of i,j is not possible. So, return ∞.

In Figure 6, for temporal distance (D,E), node D occurred at timestamp number 3. But there are no occurrences of node E also by using other intermediate occurrences of other nodes. So temporal distance (D,E) = ∞.

If there are 6 nodes, then built matrix of 6*6 and find all the values of matrix by calling a function which returns the temporal distance between source and target (i.e., it returns number of timestamp in which connection is occurred between source and target). For, Figure 6 computed temporal distance matrix values are:

Temporal Distance Matrix =

$$\begin{bmatrix} [\ 0,\ 0,\ 2,\ 2,\ -1,\ -1\ ], \\ [\ 0,\ 0,\ 2,\ 2,\ -1,\ -1\ ], \\ [\ -1,\ 1,\ 0,\ 1,\ 0,\ 0\ ], \\ [\ -1,\ 0,\ 0,\ 0,\ -1,\ -1\ ], \\ [\ -1,\ 1,\ 0,\ 1,\ 0,\ 0\ ], \\ [\ -1,\ 1,\ 0,\ 1,\ 0,\ 0\ ] \end{bmatrix} \ \ldots\ldots 1$$

**Algorithm:**

1. Input source and target, $T_{min}$ and $T_{max}$ time window
2. Time Window Equation:
$$w(Timewindow) > \frac{\sum T_{ij}(T_{min}, T_{max})}{\sum N_{ij}}$$
   Where, $\sum T_{ij}(T_{min}, T_{max})$ = Total contact time between all pair of nodes $i$, $j$ and $\sum N_{ij}$ = Total occurrences of all pair of nodes $i$ and $j$
3. Number of time frames = Tmax − Tmin / Time window
4. Initialize number of empty list equal to number of time frames. Each list shows node ids whose contact occurred in respective time frame.
5. Read the dataset and perform lookup for node contact in different time frames and generate distance matrix for each node.
6. Per contact frame, fill up the array / list with node ids in contact.
7. Compute the temporal distance as:
   a. If source and target ids are in same list, return (target's time frame number − source time frame number) as temporal distance.
   b. Otherwise, look up source and target in different time frames. If source time frame < target's time frame then return (target's time frame number − source time frame number) as temporal distance.
   c. In case repeated occurrence of source, target set $T_{min}$= last target occurred +1 timestamp and repeat steps a and b.
8. Take average values of all pairs (source, target) temporal distance.
9. Repeat steps 4,5,6 and 7 for all pairs(source, target) and generate matrix. minus one(-1) indicates no edge between pair of nodes in matrix.

In continuation of the example shown in figure 6 and successive computation of temporal distance matrix at 1, then, summation of non-negative values of matrix = 14. Now, calculate the average temporal distance value as: 300(14/ (6) (5)) =300(0.46) =140. i.e., it takes average 140 seconds for reaching from source i to destination j.

### IV Evaluation of Temporal Metrics

First network topology is generated from large real datasets using python custom made script.Python provides module called networkx , which helps to generate network topology according to dataset. For evaluation, we have downloaded the real trace data from CRAWDAD[15], and generate synthetic data by using Opportunistic Network Environment(ONE)[16] simulator. Accordingly convert datasets into common format, then calculate the Time Window size based on contact time and number of connections.

In evaluation process INFOCOM 2005, INFOCOM 2006[17] conference dataset, and Rollernet[18] real trace from INFOCOM 2009 are being used, by generating randomway point(RWP)[19] synthetic dataset using ONE simulator.

**Common dataset format for real trace data**

Table 2 below shows the common format used for datasets, and if real traces are not available in this format, then customized[1] script converts into given order and then use for calculation.

---

Table 2 Common dataset format

| Source Node ID | Destination Node ID | CONNECTION UP Time | CONNECTION DOWN Time | Occurrence Count | Inter contact Time |
|---|---|---|---|---|---|
| 1 | 3 | 51293 | 51293 | 1 | 0 |
| 1 | 3 | 60603 | 60603 | 2 | 9310 |
| 1 | 3 | 62363 | 62363 | 3 | 1760 |
| 1 | 3 | 79649 | 79649 | 4 | 17286 |

**Common format for Synthetic Data**

Customized script converts synthetic traces of RW and RWP model into connectivity report. First, input parameters for RWP model set in ONE simulation as shown in Table 3 serial number 1 and 2.

Table 3 Input parameters to ONE simulator

| 1 | Name | Synthetic Data Set (Generated using ONE Simulator) |
|---|---|---|
|  | Duration | 342915 Seconds |
|  | Participants | 98 Nodes |
|  | Mobility Model | Random Way Point |
|  | Interface range | 100 meters |
|  | Address IDs | n0 – n97 |
| 2 | Name | Synthetic Data Set (Generated using ONE Simulator) |
|  | Duration | 3096 Seconds |
|  | Participants | 63 Nodes |
|  | Mobility Model | Random Way Point |
|  | Interface range | 100 meters |
|  | Address IDs | n0 – n62 |

Then, the Figure 7 shows about connections between nodes (node1, node2), i.e., up or down (action) and at what time (simulation time).

Figure 7 Synthetic data format and generation using ONE Simulator

**Time window calculation**

About computation of time window, this has been discussed earlier in table 1 of section III. Figure 8 shows the results of time window calculation for RWP, and its value is 70.32

Figure 8 Data Set Evaluation for temporal properties

Input to script dataset filename (.dat file), Tmin(in seconds, Tmax(in seconds), time window (in seconds) and output file (.txt) which contains evaluated metrics from given dataset (.dat) file. metrics evaluated are average temporal distance, diameter, degree centrality, and betweenness and Closeness centrality and shown in Table 4.

Observations:

1. Optimal time window size varies as per number of connection between nodes, number of nodes and total duration of Tmin, Tmax.. Keeping the value < derived through script may result in overlooking connection and keeping too high will result in wastage of network resources. Hence, it preferred to take optimal in multiple of 60.

2. It found that synthetic data set values for temporal distance is poor than real trace. This is due to its characteristics moving towards center and random nature of movement. Average temporal distance values of real trace analysis enables better routing decision and it is accurate than static analysis.

3. Diameter values can be used for network density checks per time frame basis or on an average. Such, checks defines whether network is sparse or dense.

4. Different centrality values helps in identifying the important nodes. Such nodes can assist in efficient information dissemination process.

5. Referring the readings of RollerNet and RWP : It reveals that in RWP model the node movements are random and hence, number of contacts and time stamps are less, resulting in lower average temporal distance value. It is seen that most of the time the nodes are moving around centre due to which diameter, degree centrality, betweenness and closeness values are higher. These values clearly indicate the reasons (described above) behind not using the synthetic models for realistic scenarios. On the other hand, rollernet data has comparatively higher contacts, and higher number of time stamps resulting better connectivity. Therefore, for efficient information dissemination these characteristics of dataset are being used by routing engines.

## CONCLUSION AND FUTURE WORK

It reveals that the node mobility plays a vital role for efficient diffusion of information in challenged environment. And while doing so one cannot ignore to understand the movement patterns and related properties such as time order, frequency, contact duration, inter contact time, etc. These dynamic properties of connection are first analyzed and understood by using time varying matrices: temporal distance, diameter and centrality. General framework has been to design carrying capability of evaluating temporal metrics from any synthetic and real trace data. Because such frameworks help in computing number of time frames and size of time windows which in turn calculate temporal distance. These properties are very useful in designing the DTN routing protocol and understanding the dynamics of network and thereby taking forwarding or replication decision.

ACKNOWLEDGMENT

We express our sincere gratitude to the management of Ganpat University – Mehsana and Marwadi Education Foundation - Rajkot; for providing us research opportunities and their wholehearted support for such activities. Finally, our acknowledgement cannot end without thanking to the authors whose research papers helped us in making this research.

Amorim, "Density-Aware Routing in Highly Dynamic DTNs: The RollerNet Case," *IEEE Transactions on Mobile Computing*, vol. 10, no. 12, pp. 1755–1768, Dec. 2011.

[19] J. B. Tracy Camp , "A Survey of Mobility Models for Ad Hoc Network Research."

Table 4  Temporal distance, diameter, centrality and between   evaluted value for real and synthetic data set

| Dataset Details | Tmin | Tmax | Total nodes | Total no. of connections | Total no. of timestamps | time window | static distance metric | average temporal distance metric | diameter | degree centrality | betweenness centrality | closeness centrality |
|---|---|---|---|---|---|---|---|---|---|---|---|---|
| infocom 2005 (node = 41) filename = contacts.Exp3.dat timewindow = 300(233.06) | | | | | | | | | | | | |
| Day 1: 64800 - 86400 (6 hours) | 64800 | 86400 | 39 | 3411 | 72 | 300 | 1.41 | 45.35 | 4 | (15, 0.87) | (22, 0.06) | (15, 0.88) |
| Day 2: 86400 - 172800 (24 hours) | 86400 | 172800 | 40 | 8120 | 288 | 300 | 1.09 | 112.03 | 2 | (37, 1.0) | (37, 0.01) | (37, 1.0) |
| Day 3: 172800 - 259200 (24 hours) | 172800 | 259200 | 38 | 5432 | 288 | 300 | 1.1 | 134.54 | 2 | (40, 1.0) | (40, 0.0046) | (40, 1.0) |
| Day 4: 259200 - 345600 (24 hours) | 259200 | 345600 | 29 | 893 | 288 | 300 | 1.48 | 10.85 | 4 | (27, 0.86) | (11, 0.10) | (27, 0.87) |
| infocom 2006 (node = 98) filename = contacts.Exp6.dat timewindow = 3240(3129.55) | | | | | | | | | | | | |
| Day 1: 61260 - 86400 (6.98 hours) | 61260 | 86400 | 96 | 33335 | 7 | 3240 | 1.56 | 3.97 | 4 | (27, 0.81) | (85, 0.04) | (40, 0.83) |
| Day 2: 86400 - 172800 (24 hours) | 86400 | 172800 | 98 | 50697 | 26 | 3240 | 1.23 | 14.26 | 3 | (56, 0.98) | (16, 0.02) | (56, 0.98) |
| Day 3: 172800 - 259200 (24 hours) | 172800 | 259200 | 93 | 27201 | 26 | 3240 | 1.3 | 12.8 | 3 | (48, 0.95) | (51, 0.01) | (48, 0.95) |
| Day 4: 259200 - 345600 (24 hours) | 259200 | 345600 | 83 | 7642 | 26 | 3240 | 1.3 | 11.75 | 3 | (44, 0.77) | (30, 0.05) | (44, 0.80) |
| infocom 2005(trace 1)(january)(node = 9) filename = contacts.Exp1.dat timewindow = 960(886.29) | | | | | | | | | | | | |
| Day 1: 0 - 86400 (24 hours) | 0 | 86400 | 9 | 547 | 90 | 960 | 1.06 | 19.21 | 2 | (8, 1.0) | (8, 0.012) | (8, 1.0) |
| Day 2: 86400 - 172800 (24 hours) | 86400 | 172800 | 7 | 427 | 90 | 960 | 1 | 12.61 | 1 | (9, 1.0) | (9, 0.0) | (9, 1.0) |
| Day 3: 172800 - 259200 (24 hours) | 172800 | 259200 | 7 | 145 | 90 | 960 | 1.48 | 48.5 | 3 | (4, 0.83) | (1, 0.35) | (4, 0.86) |
| Day 4: 259200 - 345600 (24 hours) | 259200 | 345600 | 9 | 152 | 90 | 960 | 1.47 | 44.96 | 2 | (7, 1.0) | (7, 0.37) | (7, 1.0) |
| Iocom 2005(trace 2)(january end)(node = 12) filename = contacts.Exp2.dat timewindow = 660(573.45) | | | | | | | | | | | | |
| Day 1: 0 - 86400 (24 hours) | 0 | 86400 | 12 | 2041 | 130 | 660 | 1 | 22.5 | 1 | (12, 1.18) | (12, 0.0) | (12, 1.0) |
| Day 2: 86400 - 172800 (24 hours) | 86400 | 172800 | 12 | 634 | 130 | 660 | 1.64 | 38.83 | 3 | (10, 0.82) | (9, 0.35) | (10, 0.85) |
| Day 3: 172800 - 259200 (24 hours) | 172800 | 259200 | 11 | 703 | 130 | 660 | 1.36 | 57.61 | 2 | (9, 1.0) | (9, 0.28) | (9, 1.0) |
| Day 4: 259200 - 345600 (24 hours) | 259200 | 345600 | 11 | 582 | 130 | 660 | 1.47 | 14.89 | 3 | (10, 0.8) | (9, 0.25) | (10, 0.83) |
| Day 5: 345600 - 432000 (24 hours) | 345600 | 432000 | 3 | 195 | 130 | 660 | 1 | 2.73 | 1 | (10, 1.0) | (10, 0.0) | (10, 1.0) |
| Day 6: 432000 - 518400 (24 hours) | 432000 | 518400 | 3 | 43 | 130 | 660 | 1 | 0.96 | 1 | (10, 1.0) | (10, 0.0) | (10, 1.0) |
| RollerNet filename = contacts RollerNet.dat | | | | | | | | | | | | |
| Day 1: 1156083900 - 1156094040 | 1.156E+09 | 1156094040 | 62 | 60146 | 676 | 15 | 1.02 | 341.28 | 2 | (60, 1.0) | (60, 0.0003) | (60, 1.0) |
| Synthetic (Node = 63) filename = Synthetic_63.dat | | | | | | | | | | | | |
| Day 1: 0- 3096 | 0 | 3124 | 63 | 576 | 44 | 71 | 3.64 | 0.462012706 | 6 | (14, 0.29) | (14, 0.65) | (15, 0.42) |
| Synthetic (Node = 98) filename = Synthetic_98.dat | | | | | | | | | | | | |
| Day 1: 0- 342915 | 0 | 342936 | 98 | 4412929 | 2598 | 132 | 1.81 | 14.94 | 2 | (17, 0.99) | (17, 0.09) | (17, 0.99) |